\documentclass[11pt]{article}
\usepackage{epsfig}
\newcommand{\sss}{\vspace{.2in}}

\newcommand{\be}{\begin{equation}}
\newcommand{\ee}{\end{equation}}
\newcommand{\bea}{\begin{eqnarray}}
\newcommand{\eea}{\end{eqnarray}}
\newcommand{\sn}{{\rm sn}}
\newcommand{\cn}{{\rm cn}}
\newcommand{\dn}{{\rm dn}}
\newcommand{\sech}{{\rm sech}}

\begin{document}
\sss
\sss
\begin{center}
{\Large {\Large \bf A QES Band-Structure Problem in One Dimension}}
\end{center}
\vspace{.5in}
\begin{center}
{\large{\bf
   \mbox{Avinash Khare}\footnote{khare@iopb.res.in}}}
\end{center}
\vspace{.2in}
\noindent
Institute of Physics, Sachivalaya Marg, Bhubaneswar 751005, India\\
\sss
\begin{abstract}
I show that the potential
$$V(x,m) = \big [\frac{b^2}{4}-m(1-m)a(a+1) \big ]\frac{\sn^2 (x,m)}{\dn^2 (x,m)} -b(a+\frac{1}{2}) \frac{\cn (x,m)}{\dn^2 (x,m)}$$
constitutes a QES band-structure problem in one dimension. In
particular, I show that for any positive integral or half-integral $a$, $2a+1$
band edge eigenvalues and eigenfunctions can be obtained analytically. 
In the limit of m going to 0 or 1, I recover the
well known results for the QES double sine-Gordon or double sinh-Gordon
equations respectively. As a by product, I also obtain the bound state
eigenvalues and eigenfunctions of the potential
$$V(x) = \big [\frac{\beta^2}{4}-a(a+1) \big ] \sech^2 x +\beta(a+\frac{1}{2})\sech x\tanh x$$
in case $a$ is any positive integer or half-integer.
\end{abstract}
\newpage

\sss

In last few years, the quasi-exactly solvable (QES) problems have received
wide attention in the literature \cite{1,2}. In these cases, the corresponding
orthogonal polynomials satisfy a three-term recursion relation. Further, the
Hamiltonian (or its gauged transform) can be written in terms of at most the
quadratic combination of the generators of the $SL(2,R)$ group. Two of the
celebrated QES potentials are the double sine-Gordon (DSG) potential
\be\label{1}
V(x) = \frac{b^2}{4} \sin^2 x -b(a+\frac{1}{2})\cos x~,
\ee
and the double sinh-Gordon (DSHG) potential \cite{3}
\be\label{2}
V(x) = \frac{b^2}{4} \sinh^2 x -b(a+\frac{1}{2})\cosh x~.
\ee
In both these cases, the eigenvalues and eigenfunctions for $2a+1$ levels are
analytically known in case $a$ is any positive integer or half-integer.
Further, the properties of the corresponding orthogonal polynomials have been
studied in some detail \cite{4}.

The question that I would like to raise and answer in this note is the
following: Instead of solving both the DSG and DSHG problems, is it not
possible to solve one `elliptic' problem of which the potentials (1) and (2)
constitute special cases? In this note I show that the answer to this question
is yes. In particular, I solve the Schr\"odinger equation for the periodic
potential
\be\label{3}
V(x,m) = \big [\frac{b^2}{4}-m(1-m)a(a+1) \big ]
\frac{\sn^2 (x,m)}{\dn^2 (x,m)}
-b(a+\frac{1}{2}) \frac{\cn (x,m)}{\dn^2 (x,m)}~,
\ee
and show that it is a QES band-structure problem, i.e. $2a+1$ band edge
eigenvalues and eigenfunctions can be analytically obtained in case $a$ is
any positive integer or half-integer. In particular, explicit expressions
for band-edge eigenstates are given in case $a=0,1/2,1,3/2,2$. Not
surprisingly, for $m=0,1$ we recover the well known results for the DSG and DSHG
potentials (1) and (2) respectively.
Here $\cn (x,m),\sn(x,m)$ are the Jacobi elliptic functions of real elliptic
modulus parameter m ($0 \le m \le 1$) with period $4K(m)$ while $\dn(x,m)$
has period $2K(m)$. For simplicity, now onward, we will not explicitly
display the modulus parameter m as an argument of Jacobi elliptic functions
\cite{5}.

I also show that in case $a$ is any positive integer or half-integer
then the (gauged) Hamiltonian can be written in terms of at most the
quadratic generators of the $SL(2,R)$ group. Further,  
the associated orthogonal polynomials satisfy three-term 
recursion relation. Finally, as a by product, I also obtain
the bound state eigenvalues and eigenfunctions of the well known exactly
solvable potential \cite{6}
\be\label{4}
V(x) = \big [\frac{\beta^2}{4}-a(a+1) \big ] \sech^2 x
-\beta(a+\frac{1}{2})~\sech x \tanh x~,
\ee
in case $a$ is any positive integer or half-integer.

We start from the Schr\"odinger equation ($\hbar = 2m =1$)
\be\label{5}
\frac{d^2 \psi(x)}{dx^2} +[E-V(x)]\psi(x) = 0~,
\ee
where $V(x)$ is as given by eq. (\ref{3}). Note that the potential (\ref{3})
is of period $4K(m)$, where $K(m)$ denotes the complete elliptic integral
of the first kind. In fact, the potential (\ref{3}) is also invariant
under $x \rightarrow x+2K(m)$ provided we also let $b \rightarrow -b$. Further,
in the limit m going to 0 or 1, the potential (\ref{3}) reduces to the DSG
or DSHG potentials (\ref{1}) and (\ref{2}) respectively.
In this note, we are interested in obtaining the band edge eigenvalues and
eigenfunctions which if arranged in order of increasing energy
$E_0 < E_1 \le E_2 < E_3 \le E_4$...are of period $4K, 8K, 8K, 4K, 4K,$...
with the corresponding number of wave function nodes in the interval $4K$
being 0,1,1,2,2,... .

We substitute
\be\label{6}
\psi = \exp \bigg [-\frac{b}{2\sqrt{m(1-m)}}
tan^{-1} (\sqrt{\frac{m}{1-m}} \cn \, x) \bigg ]y~,
\ee
in the Schr\"odinger equation (\ref{5}) yielding\
\be\label{7}
y''(x) -b \frac{\sn \,x}{\dn \,x} y'(x)
+\big [E + ab \frac{\cn \, x}{\dn^2 \, x}
+m(1-m)a(a+1)\frac{\sn^2 \,x}{\dn^2 \, x} \big ]y(x) = 0~.
\ee
On further substituting
\be\label{8}
y(x) = (\dn \, x)^{-a} u(x)~,
\ee
it is easily shown that $u(x)$ satisfies the equation
\bea\label{9}
u''(x) &+& \big [2am \frac{\sn \, x \cn \, x}{\dn \, x}
-b\frac{\sn \, x}{\dn \, x} \big ]u'(x) \nonumber \\
       &+& \big [E+am+ab \cn \, x +ma(a-1) \sn^2 \, x \big ]~u(x) = 0~.
\eea

We now want to show that irrespective of the value of b,
eq. (\ref{9}) is a QES case. In particular, we wish
to show that if a is a positive integer (half-integer) than eq. (\ref{9})
admits $2a+1$ algebraic
solutions of period $4K$ ($8K$). For the special case of $b=0$ this is
of course well known since in that case the potential (\ref{3}) essentially
reduces to the Lam\'{e} potential.

Let us start from eq. (\ref{9}) and substitute
\be\label{10}
\sn \, x = \sin \theta~, \ u(x) \equiv z(\theta)~.
\ee
Then $z(\theta)$ satisfies
\bea\label{11}
(1-m\sin^2 \theta)z''(\theta)
+\big [(2a-1)m\cos \theta \sin \theta
-b\sin \theta \big ]z'(\theta) \nonumber \\
+\big [E+am+ab \cos \theta +ma(a-1)\sin^2 \theta \big ]z(\theta) = 0~.
\eea
On further substituting $\cos \theta =t$, one finds that $z(t)$ satisfies
\bea\label{12}
\big [mt^4+(1-2m)t^2-(1-m) \big ]z''(t) \nonumber \\
+\big [2m(1-a)t^3+bt^2+(2ma-2m+1)t-b \big ]z'(t) \nonumber \\
+ \big [ma(a-1)t^2-abt-E-ma^2 \big ]z(t)=0~.
\eea
It is now straight forward to check that eq. (\ref{12}) can be written as a
quadratic combination of the operators
\be\label{13}
J_n^{+} = t^2\frac{d}{dt} -nt~, \ J_n^{0} = t\frac{d}{dt} -\frac{n}{2}~, \
J_n^{-} = \frac{d}{dt}~,
\ee
where $J_n's$ are the generators of the non-compact Lie group $SL(2,R)$,
provided $a=n$. In particular, for $a=n$, eq. (\ref{12}) can be written as
\be\label{14}
\bigg [ mJ_n^{+}J_n^{+} +(1-2m)J_n^{0}J_n^{0} -(1-m)J_n^{-}J_n^{-}
+nJ_n^{0}+b(J_n^{+}-J_n^{-})+\lambda \bigg ]z(t) = 0~, 
\ee
where $\lambda = -(E+\frac{mn^2}{2} -\frac{n^2}{4})$. Thus, if $a$ is a
positive integer $n$, then the three generators $J_n^{\pm,0}$ form a
representation of dimension $n+1$ of the group $SL(2,R)$.

Similarly, on substituting $z(\theta) = \sin \theta w(t=\cos \theta)$
in eq. (\ref{11}), it is easily shown that the resulting equation can again
be written as a quadratic combination of the generators (\ref{13})
provided $a = n+1$. Thus if $a$ is a positive integer, then
the three generators $J_n^{\pm,0}$ form a representation of dimension $n$
of the group $SL(2,R)$. In this way, we have shown that
when $a$ is a positive integer $n$, then one will have $n+1$ solutions of
the type $F_n (t)$ and $n$ solutions of the type $\sin \theta F_{n-1} (t)$.

Let us now turn to the case of half-integral $a$. On substituting
\be\label{15}
t = \frac{1+\cos \theta}{2}~, \ z(\theta) = t^{1/2} w(t)~,
\ee
in eq. (\ref{11}) it is easily shown that the resulting equation can be
written as a quadratic combination of the generators (\ref{13})
provided $a=n+\frac{1}{2}$. A similar conclusion is also reached in case we
substitute
\be\label{16}
t = \frac{1-\cos \theta}{2}~, \ z(\theta) = t^{1/2} w(t)~,
\ee
in eq. (\ref{11}). Thus, in the half-integral case (i.e. $a=n+\frac{1}{2}$),
we have shown that one has $n+1$ solutions of the type
$\cos \frac{\theta}{2} F_n (t)$ and $n+1$ solutions of the type
$\sin \frac{\theta}{2} F_n (t)$.

As an illustration, we now give explicit solutions for few values of $a$.
In particular, we specify the eigenvalue E and eigenfunction $u(x)$ with
$\psi$ being related to $u$ by eqs. (\ref{6}) and (\ref{8}).

a =0:

\be\label{17}
E =0~, \ u(x) = constant~.
\ee

$a=\frac{1}{2}$:

\be\label{18}
E = \frac{1-2m\mp 2b}{4}~, \ u(x) = \sqrt{1\pm \cn \, x}~.
\ee

a=1:

\be\label{19}
E = 1-2m~, \ u(x) = \sn \, x~,
\ee
\be\label{20}
E= \frac{1-2m\pm\sqrt{1+4b^2}}{2}~, \ u(x) = b - (E+m) \cn \, x~.
\ee

$a=\frac{3}{2}$:

\bea\label{21}
E &=& \frac{5-10m-2b}{4} \pm \sqrt{1-m(1-m)+(1-2m)b+b^2}~, \nonumber \\
u(x) &=& (\alpha + \beta \cn \, x)\sqrt{1+\cn \, x}~,
\eea
\bea\label{22}
E &=& \frac{5-10m+2b}{4} \pm \sqrt{1-m(1-m)-(1-2m)b+b^2}~, \nonumber \\
u(x) &=& (\alpha_1 + \beta_1 \cn \, x)\sqrt{1-\cn \, x}~.
\eea

a = 2:

\be\label{23}
E = \frac{5-10m}{2} \pm \sqrt{9+4b^2}~,
\ u(x) = \sn \,x (\alpha_2 +\beta_2 \cn \, x)~,
\ee
\bea\label{24}
x^3 &+& 2(2m-1)x^2 -(4b^2+3)x +8(1-2m)b^2 = 0~, \ E = x+1-2m~, \nonumber \\
u(x) &=& \alpha_3 +\beta_3 \cn \, x +\delta_3 \cn^2 \, x~.
\eea
In the above equations, the constants $\alpha, \beta$ etc. are easily
determined.

As expected, for $m=0,1$, these eigenvalues and eigenfunctions agree with
the well known results for the DSG and DSHG potentials (\ref{1}) and
(\ref{2}) respectively.

For the special case of $m=\frac{1}{2}$, the cubic eq. (\ref{24})
is easily solved yielding $E=0, \pm \sqrt{4b^2+3}$. It is amusing to notice
that for $m=\frac{1}{2}$, the eigenvalues given in eqs. (\ref{17}) to
(\ref{24}) are symmetric about $E=0$. We do not know if there is any deeper
reason for it.

As is well known, whenever one obtains QES solutions, the associated
orthogonal polynomials satisfy a three-term recursion relation and this is 
also true in the present case. In this context we recall the detailed work
of Finkel et al. \cite{fin}. In their language, our periodic problem 
as given by eq. (12) corresponds to the case (7) in their notation 
(see their eqs. (8) and (20)) and hence on running through the steps 
given there it follows that in our case the orthogonal polynomials
satisfy a three-term recursion relation.

Before ending this note we show that as a by product, we also obtain the
band edge eigenvalues and eigenfunctions of the periodic potential
\be\label{28}
V(x,m) = \big [\frac{\beta^2}{4}-ma(a+1) \big ]\cn^2 (x,m)
+\beta(a+\frac{1}{2}) \sn (x,m) \dn (x,m)~,
\ee
in case $a$ is either an integer or a half-integer.
Since in the limit $m \rightarrow 1$, this potential goes over to the
exactly solvable potential (\ref{4}), hence we also obtain the
bound state eigenvalues and eigenfunctions of the potential (\ref{4})
in that case. The proof is rather simple. Since under $x \rightarrow x+K(m)$
\be\label{29}
\sn \, x \rightarrow \frac{\cn \, x}{\dn \, x}~, \ \cn \, x \rightarrow
-\sqrt{1-m} \frac{\sn \, x}{\dn \, x}~,
\ \dn \, x \rightarrow \frac{\sqrt{1-m}}{\dn \, x}~,
\ee
hence, under this transformation the potential (\ref{3}) goes over to the
potential (\ref{28}) with $\beta = -\frac{b}{\sqrt{1-m}}$. Thus the
band-edge eigenvalues and eigenfunctions of the potential (\ref{28}) are
immediately obtained from those of the potential (\ref{3}) by making the
substitution
\be\label{30}
b \rightarrow -\sqrt{1-m} \, \beta~,
\ \sn \, x \rightarrow -\frac{\cn \, x}{\dn \, x}~, \
\cn \, x \rightarrow \sqrt{1-m}\frac{\sn \,x}{\dn \, x}~,
\ \dn \, x \rightarrow \frac{\sqrt{1-m}}{\dn \, x}~.
\ee
It is amusing to note that while the eigenvalues so obtained are in general
$\beta$
dependent,  as $m \rightarrow 1$, this $\beta$ dependence completely
disappears from the eigenvalues (note, one is replacing $b$ by
$\sqrt{1-m} \, \beta$), as it should since it is well known \cite{6}
that
the eigenvalues of the potential (\ref{4}) are $\beta$ independent.
Note, however, that the corresponding eigenfunctions are still
$\beta$-dependent.

Summarizing, it is nice to know that the QES solutions of the `elliptic'
potential (\ref{3}) automatically give us the well known QES solutions of
the DSG and DSHG potentials (\ref{1}) and (\ref{2}) respectively. Further,
at the same time, it also gives us the bound state eigenvalues and
eigenfunctions of the well
known shape invariant potential (\ref{4}) \cite {6}
in case $a$ is positive integer or half-integer. It will be interesting
to explore
if other QES problems can also be dealt with in this unified manner.

\end{document}